\documentclass[prl,twocolumn,showpacs,preprintnumbers,amsmath,amssymb]{revtex4}
\usepackage{graphicx}
\usepackage{dcolumn}
\usepackage{bm}
\begin{document}

\title{Kondo Quantum Dots and the Novel Kondo-doublet interaction.}
\author{J. Simonin}
\affiliation{Centro At\'{o}mico Bariloche and Instituto Balseiro, \\
8400 S.C. de Bariloche, R\'{i}o Negro,  Argentina}
\date{today}

\begin{abstract}
We analyze the interactions between two Kondo Quantum Dots
connected to a  Rashba-active Quantum Wire. We find that the
Kondo-doublet interaction, at an inter-dot distance of the order
of the wire Fermi length, is over an order of magnitude greater
than the RKKY interaction. The effects induced on the
Kondo-doublet interaction by the wire spin-orbit coupling can be
used to control the Quantum Dots spin-spin correlation. These results
imply that the widely used assumption that the RKKY is the
dominant interaction between Anderson impurities must be revised.
\end{abstract}
\maketitle

\textit{Introduction}. Kondo Quantum Dots made in semiconductor
heterostructures sparked a growing experimental and theoretical
research in the area \cite{gold,craig,vincenzo}. In those systems
a correlated electronic state localized in a quantum dot (QD)
interacts with a low dimensional electron gas via an hybridization
term. They belong to the very rich family of physical systems
that are described by the Anderson-Impurity Hamiltonian \cite{hewson}.
The Kondo Quantum Dots regime, in which a single electron is allowed in the QD top most
populated orbital due to the Coulomb blockade effect, has
attracted considerable interest due to the possibility of using the
QD spin as a quantum bit \cite{loss,chuang}. These  systems have a
relatively high Kondo ($\delta_K \simeq 1 K$) to Fermi ($E_F$)
energy ratio, corresponding to a strong Kondo coupling of the
dot-band system \cite{sato1}. The two Kondo QD problem is crucial for understanding how QDs interact, and how these interactions can be controlled. This has been the subject of
several recent letters \cite{tamura,carlos1,pascal1,glazm,carlos2}. 

A common denominator of those letters is that they assume that the
correlation between the QD's  is given by the
RKKY interaction \cite{hewson,kittel}. In the Kondo regime the RKKY
is the strongest of the different fourth order interactions present
in the two-QD Anderson Hamiltonian \cite{falicov,cesar,indisw}.
But it was recently found in Ref.\cite{jsfull} that the dominant
interaction in the Kondo regime is the non-perturbative second order
Kondo-doublet process.

The Kondo-doublet owes its strong Kondo-like correlation energy to
the resonance of one of the sets of configurations
that form its screening cloud. The set with one spin up electron
in each QD plus a spin down hole in the band interact via the
annihilation of the hole in any of the two QDs. Thus, the
``connectivity" of this set is enhanced by the interference of
these two paths. As the enhanced set is one of parallel QD-spins, the Kondo-doublet generates a strong ferromagnetic correlation between the QDs.

Therefore, two Kondo QDs connected to a Rashba active
quantum wire (QW) is an ideal system to check the predictions of Ref.\cite{jsfull}.
 QDs-QW systems are experimentally accessible \cite{sasaki,sato1}. A simple tool for fine tuning the interactions between the QDs is a tunable Rashba \cite{rashba} spin-orbit coupling (RSO) of the band states \cite{molen}, that makes the spin of the band excitations to precess. Such a system is also very important for the development of nanoscopic quantum electronics \cite{loss}. In the following we analyze the effects of the RSO interaction on the Kondo-doublet for a two-QD-QW setup. We compare our results with that of Imamura \textit{et al}
\cite{imamura} for the RKKY interaction. We find that the Kondo-doublet interaction is over an order of magnitude greater than the RKKY interaction.

\begin{figure}[h]
\includegraphics[width=\columnwidth]{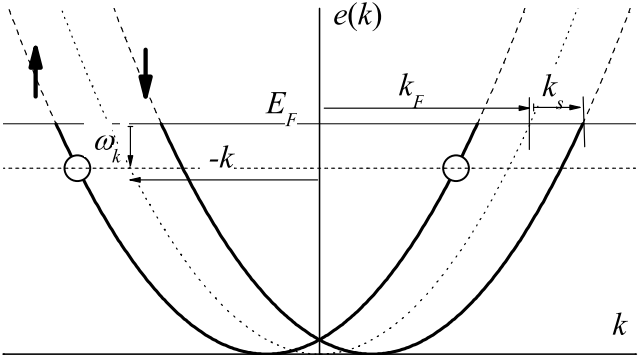}
\caption{Quantum wire energy bands. The spin-degeneracy is removed
by the Rashba spin-orbit coupling. The circles on the spin up band
mark the two spin down holes of energy $w_k$. } \label{fig1}
\end{figure}

\textit{Model}. For a quantum wire with RSO it is convenient to choose the
spin quantification axis perpendicular to both the wire and the
effective RSO electric field, conserving $S_z$ as a good quantum
number. Thus, with the wire along the $x$-axis and the RSO
electric field applied in the $y$-axis the Hamiltonian for
the conduction electrons is given by $ H_b=-\frac{\hbar^2}{2m}\ \partial_x^2- i\ \alpha \ \partial_x\ \sigma_z $ , where $\alpha$ is the variable RSO coupling  and $\sigma_z$ the
Pauli matrix. The spin-degeneracy of the band states is removed.
The spin-up (down) $k$-states energy is now $e_{k
\uparrow(\downarrow)}= \frac{\hbar^2}{2m}(k\pm k_s)^2-\Delta_s$,
where $k_s= \frac{m}{\hbar^2}\ \alpha $. The effect of the last
term ($\Delta_s = \frac{\hbar^2}{2m}\ k_s^2$) is to lower the
bottom of the band, increasing the electron density in the wire if
the chemical potential is maintained fixed. In the following we
disregard this term, working at fixed QW electron density, Fig.
\ref{fig1}. If necessary, the effects of this term  can be
included in the final results by renormalizing the Fermi energy and
wave-vector  ( $E_F \mapsto E_F + \Delta_s$ and $k_F \mapsto
\sqrt{k_F^2+k_s^2}$ ) \cite{imamura}. We concentrate our work in
the phase effects generated by the RSO.

The two-QD-QW Anderson Hamiltonian is the sum of the band, hybridization ($H_V$),
 and correlated QDs (with the QDs at $R_j=\pm R/2$) Hamiltonians
\begin{eqnarray}\label{hamil}
H=\sum_{k \sigma} e_{k \sigma} \ c^\dag_{k\sigma} c_{k\sigma} +
\textbf{v} \sum_{j k \sigma}(e^{i\; k R_j} \ \
d^\dag_{j\sigma} c_{k\sigma}+ h.c. )\nonumber \\
 - E_d \sum_{j \sigma} d^\dag_{j\sigma} d_{j\sigma} +
 U \sum_j d^\dag_{j\uparrow} d_{j\uparrow}d^\dag_{j\downarrow}d_{j\downarrow}\ , \ \ \ \ \ \ \ \
\end{eqnarray}
where the fermion operator $c_{k\sigma}$($d_{j\sigma}$) act on the
conduction band $k$-state (on the QD at $R_j$)
and $\textbf{v}=V/\sqrt{L}$ is the hybridization  divided by the
square root of the wire length. We renormalize the vacuum (denoted
by $|F\rangle$) to be the conduction band filled up to the Fermi
energy and we make an electron-hole transformation for band states
below the Fermi level: $b^\dag_{k\overline{\sigma}}\equiv
c_{k\sigma}$. Single state energies are referred to the Fermi
energy. In the Kondo limit, the case analyzed in this work, the QD
level is well below the Fermi energy ($-E_d \ll 0$), and it can
not be doubly occupied due to the strong Coulomb repulsion ($U \gg E_d$). In this regime the two relevant Hamiltonian parameters are the effective Kondo coupling $J_n=\rho_o V^2/E_d$ ($\rho_o$ being the density of band states at the Fermi level) and
the inter-dot distance $R$. The single QD Kondo energy is given by
$\delta_K = E_F\ \exp{(-1/2 J_n)}$. We use in the text a ``ket"
notation for the QDs-band configurations, the first symbol
indicates the status of the left QD (the one at $x=-R/2$) and the
second one the status of the QD on the right, e.g. $\ | 0 0
\rangle \equiv|F\rangle$, $\ | 0 \! \uparrow \rangle \equiv
d^\dag_{R\uparrow}|F\rangle$, $|\!\! \downarrow \uparrow \rangle
\equiv d^\dag_{L\downarrow} d^\dag_{R\uparrow} |F\rangle$,
 $|\!\!\downarrow \uparrow, h_{k\sigma}\rangle \equiv b^\dag_{k\sigma}d^\dag_{L\downarrow}
 d^\dag_{R\uparrow} |F\rangle$, etc..

With the chosen axis the non-resonant set (composed by the
$|\!\!\uparrow \downarrow, h_{k\uparrow}\rangle$ and $|\!\!
\downarrow \uparrow, h_{k\uparrow} \rangle$ configurations) remains
so due to $S_z$ conservation. The effects of the RSO
appear in the  resonant set-vertex states connections (matrix elements). Due to the mirror symmetry breaking produced by the RSO the even and odd doublets are mixed. To trace these effects we select two components of the set, the ones with a hole
of energy $w_k$. They correspond to the removal of the
$k_-=-k-k_s$ ($|\!\!\uparrow \uparrow, h_{k_-\downarrow}\rangle$)
or the $k_+=k-k_s$ ($|\!\!\uparrow \uparrow,
h_{k_+\downarrow}\rangle$) electron from the spin-up band, see
Fig. \ref{fig1}. Applying $H_V$ to the vertex state
($|A_\uparrow\rangle=|0\! \uparrow\rangle-|\!\!\uparrow0\rangle$)
of the odd doublet the following combination is generated
\begin{equation}\label{bc1}
 -(e^{i \varphi_+}+e^{-i \varphi_+})|\!\!\uparrow \uparrow,
 h_{k_+\downarrow}\rangle-(e^{i \varphi_-}+e^{-i
\varphi_-})|\!\!\uparrow \uparrow, h_{k_-\downarrow}\rangle \ ,
\end{equation}
where $\varphi_\pm = k_\pm R/2$. The contribution of these configurations to the set connectivity is obtained by closing the Kondo-doublet path (in units of $4 V^2/L$),
\begin{equation}
\langle A_\uparrow|H_V^2|A_\uparrow \rangle_{\uparrow\uparrow}= 1 + \cos{(k R)} \cos{(k_s R)} \ ,
\end{equation}
the first term on the \textit{r.h.s.} ($1$) comes from the hole being
annihilated at the same QD where it was generated and the second
term corresponds to the annihilation of the hole at the other QD.
This last term includes the additional phase factor generated
by the RSO. But now, due to the RSO symmetry breaking the configuration of (Eq.(\ref{bc1})) is also connected with the  even doublet vertex state ( $|S_\uparrow\rangle=|0\! \uparrow\rangle+|\!\!\uparrow0\rangle$), given that
\begin{equation}
\langle S_\uparrow|H_V^2|A_\uparrow \rangle_{\uparrow\uparrow}= - \ i \ \cos{(k R)} \sin{(k_s R)} \ .
\end{equation}
A straightforward analysis shows that the proper vertex states when the
RSO is active is the twisted ``odd" vertex
\begin{equation}\label{two}
|\widetilde{A_\uparrow} \rangle= \cos{\varphi_s}|A_\uparrow \rangle -
i \sin{\varphi_s}|S_\uparrow \rangle = e^{-i \varphi_s}|0\!
\uparrow\rangle - e^{i \varphi_s}|\!\!\uparrow0\rangle \ ,
\end{equation}
for the generation of a ``twisted-odd" doublet, and the twisted ``even" vertex
\begin{equation}\label{twe}
|\widetilde{S_\uparrow} \rangle= \cos{\varphi_s}|S_\uparrow \rangle -
i  \sin{\varphi_s}|A_\uparrow \rangle = e^{-i \varphi_s}|0\!
\uparrow\rangle + e^{i \varphi_s}|\!\!\uparrow0\rangle \ ,
\end{equation}
for an orthogonal ``twisted-even" doublet, where $\varphi_s = k_s R/2$. Thus, for the ``twisted odd" doublet the $w_k$ components of the Kondo-doublet cloud are
\begin{equation}\label{bc2}
 |\!\!\uparrow\uparrow,w_{k\downarrow} \rangle=(e^{i \varphi_k}+
 e^{-i \varphi_k}) (|\!\!\uparrow \uparrow,
h_{k_+\downarrow}\rangle + |\!\!\uparrow \uparrow,
h_{k_-\downarrow}\rangle) \ ,
\end{equation}
in the resonant set, and
\begin{eqnarray}\label{bu2}
|\sigma\overline{\sigma},w_{k\uparrow} \rangle = e^{i (\varphi_k-2\varphi_s)}
|\!\!\downarrow \uparrow, h_{\overline{k}_-\uparrow}\rangle+e^{-i (\varphi_k+2\varphi_s)}
|\!\!\downarrow \uparrow, h_{\overline{k}_+\uparrow}\rangle \nonumber \\
+ e^{-i (\varphi_k-2\varphi_s)}|\!\!\uparrow \downarrow,
h_{\overline{k}_-\uparrow}\rangle+e^{i
(\varphi_k+2\varphi_s)}|\!\!\uparrow \downarrow,
h_{\overline{k}_+\uparrow}\rangle \ ,\ \ \ \ \ \
\end{eqnarray}
in the non-resonant set. In the equations above $\varphi_k = k
R/2$, and $\overline{k}_\pm=\pm k + k_s $ are the $k$-wave vectors
of the spin up holes of energy $w_k$. For
$|\widetilde{A_\uparrow} \rangle$ the odd doublet connectivity factor
for the resonant set path is recovered, given that $\langle
\widetilde{A_\uparrow}|H_V^2|\widetilde{A_\uparrow}
\rangle_{\uparrow\uparrow}= 1 + \cos{(k R)}$. Note that the effect of the
twisted vertex is to remove the RSO phases from the resonant set
(compare Eq.(\ref{bc2}) with Eq.(\ref{bc1})); the RSO phase
effects accumulate in the non-resonant set, Eq.(\ref{bu2}).

Kondo-like interactions can not be analyzed by standard perturbative methods (\cite{kittel} pag. 155, \cite{hewson} pags. 53, 149). Therefore, adding over all the possible values of the hole
energy, we analyze the properties of the ``twisted odd" doublet by
means of the following variational wave function
\begin{eqnarray}\label{twodd}
|\widetilde{D_{o\uparrow}}\rangle= |\widetilde{A_\uparrow}\rangle + \textbf{v} \sum_{w_k}\  Z(w_k)
(|\!\!\uparrow\uparrow,w_{k\downarrow} \rangle+|\sigma\overline{\sigma},w_{k\uparrow} \rangle)\ ,
\end{eqnarray}
where both cloud sets must be included to account for the synergy of Kondo structures.
Analytical minimization \cite{varma,jsfull} gives the energy  of
the ``twisted odd" doublet ($E_o = -2E_d-\delta_o(R)$), and the
variational amplitude of the configurations in its cloud
($Z(w)=1/(\delta_o+w)$), where $\delta_o$ is the bare (no RSO
coupling) odd doublet correlation energy \cite{jsfull}
\begin{equation}\label{dodd}
\delta_o(R)=E_F\ \exp{ \frac{-1}{[2 \pm C_Q(\delta_o, R)]\ J_n} }\
,
\end{equation}
where the minus sign is for the even doublet correlation energy
($\delta_e$) and the hole coherence factor $C_Q(\delta,R)$ (
$\equiv (\Sigma_w \cos{(k R)}\ Z(w))/(\Sigma_w Z(w))$ ) is given
by
\begin{equation}\label{cqh}
C_Q = \frac{\cos{(u)} [ \text{Ci}(u)-\text{Ci}(t)] + \sin{(u)}[
\text{Si}(u)-\text{Si}(t)]}{\ln{(1+E_F/\delta)}}\ ,
\end{equation}
where  $t = k_F R$ and $u = k_F R(1+\delta/E_F)$, $\text{Ci}$
($\text{Si}$) is the CosIntegral (SinIntegral)
function.

\begin{figure}[h]
\includegraphics[width=\columnwidth]{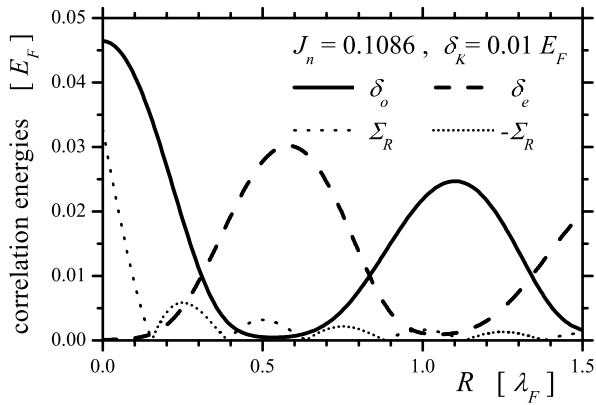}
\caption{Correlation energy of the Kondo-doublets and the RKKY
configurations as a function of the distance between the QDs. At
$R=0$ the odd-doublet interaction is just a fifty per cent greater
than the RKKY value. At $R\simeq \lambda_F$ the second maximum of the odd-doublet is
already $15$ times greater than the third ferromagnetic maximum of
the RKKY. } \label{fig2}
\end{figure}

Therefore, as in the RKKY case, the band RSO coupling
does not modify the strength of the Kondo-doublet interaction. In
Fig.\ref{fig2} we plot $\delta_o$, $\delta_e$, and the RKKY
 energies ($\pm \Sigma_R$, with $\Sigma_R(R) = 4\ln{2}\ J_n^2\ E_F \ [1 -(2/\pi)\ \text{Si}(2 k_F R)]$  \cite{litvi, imamura, jsfull}) as function of the inter-dot distance $R$.
Notice that the RKKY decays faster than the Kondo-doublet
interaction, at $R\simeq 2 \lambda_F$ the third maximum of the
odd-doublet is $25$ times greater than the fifth ferromagnetic
maximum of the RKKY.

But the internal structure of the doublets has been ``twisted" by
the RSO phase effects. These changes are reflected in the QDs spin-spin
correlation. For the ``twisted odd" doublet it is given by
\begin{equation}\label{slsr}
\frac{\langle \widetilde{D_{o\sigma}}| S_L.S_R |\widetilde{D_{o\sigma}}\rangle}
{\langle \widetilde{D_{o\sigma}}|\widetilde{D_{o\sigma}}\rangle} = \pm
\frac{(1+2\cos{2 k_s R})D_Q(R)}{4(2 \pm D_Q(R))} \ ,
\end{equation}
where $D_Q (R) \equiv (\Sigma_w \cos{(k R)}\  Z(w)^2)/(\Sigma_w
Z(w)^2)$ and the lower sign holds for the ``twisted even" doublet.
\begin{figure}[h]
\includegraphics[width=\columnwidth]{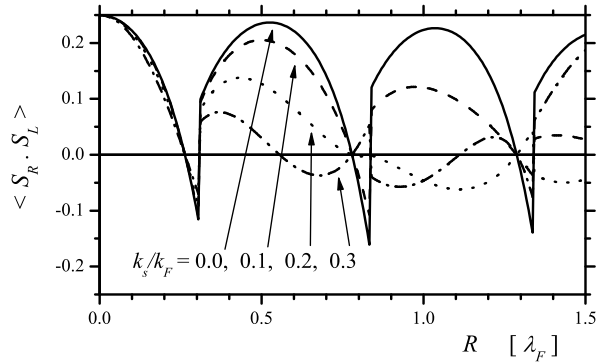}
\caption{QDs-spin correlation for the dominant doublet for
different values of the RSO coupling. For $R\simeq \lambda_F$ a
moderate value of the effective  RSO coupling turns the parallel
alignment of the QDs spin into a slightly antiparallel one. }
\label{fig3}
\end{figure}

\begin{figure}[h]
\includegraphics[width=\columnwidth]{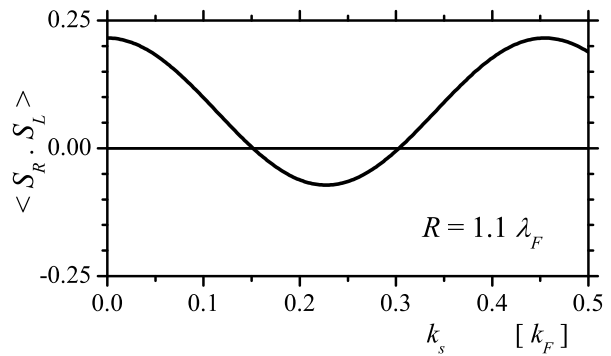}
\caption{QDs-spin correlation for the dominant ``odd" doublet at
$R = 1.1 \lambda_F$ as a function of the RSO coupling.}
\label{fig4}
\end{figure}

In Fig.\ref{fig3} we plot $\langle S_L.S_R\rangle$ for the
dominant doublet (\textit{i.e.} that with the highest $\delta(R)$) for different values of the effective RSO coupling $k_s$. As the twisting angle is given by $ 2 k_s R $ this
effect is greater the greater $R$. This amplification property of
$R$ on the RSO effects must be balanced, in an experimental setup,
with the fact that the correlation between the QDs decays due to
the increasing decoherence of the hole packet that generates the
Kondo-doublet interaction. A good experimental compromise is
$R\simeq \lambda_F$, about a few dozen nanometers for these
heterostructures. For this distance a moderate transversal electric
field can turn the parallel spin alignment typical of the dominant
Kondo-doublet into a slightly antiparallel one. In Fig.\ref{fig4} we show
$\langle S_L.S_R \rangle$ for the dominant ``odd" doublet at
$R = 1.1 \lambda_F$ as a function of the RSO coupling.

We compare the effects of the RSO on the Kondo-doublets with those on the RKKY. 
We recalculate the results of Ref.\cite{imamura}, which can be
summarized as follows: the strength of the RKKY does not change
(apart from the Fermi energy renormalization discussed previously),
but the structure of the RKKY configurations is ``twisted". Two of
the eigenvectors of the Imamura RKKY-RSO Hamiltonian
(Ref.\cite{imamura}, Eq.(19) plus a $-\Sigma_R/2$ term) are, in
our axis setup, the $|S_z|=1$ components of the zero-RSO
ferromagnetic triplet ($|\!\!\uparrow\uparrow \rangle$ and
$|\!\!\downarrow\downarrow \rangle$) with a correlation energy
gain $\Sigma_R$, and the other two are a twisted mixture of the two $S_z=0$
configurations, the odd ferromagnetic $|FM0 \rangle$) and the even
antiferromagnetic $|AF0\rangle$ ones ($(|\!\!\downarrow\uparrow
\rangle \pm |\!\!\uparrow\downarrow \rangle)/\sqrt{2}$). The
twisted ``ferromagnet", with a correlation energy gain $\Sigma_R$, is
given by
\begin{eqnarray}\label{twfm}
|\widetilde{FM0}\rangle= \cos{4\varphi_s}|FM0\rangle + i  \sin{4\varphi_s}|AF0\rangle = \nonumber \\
e^{i 2 \varphi_s}|\!\!\downarrow\uparrow\rangle + e^{-i 2 \varphi_s}|\!\!\uparrow\downarrow\rangle\
\ \ \ \ \ \ \ \ \ ,
\end{eqnarray}
and the twisted ``antiferromagnet", of energy gain $-\Sigma_R$, by
\begin{equation}
|\widetilde{AF0}\rangle= \cos{4\varphi_s}|AF0\rangle + i  \sin{4\varphi_s}|FM0\rangle \ .
\end{equation}
Direct evaluation of $\langle S_R.S_L\rangle$ for the twisted
ferromagnet gives $(2 \cos{2 k_s R}-1)/4 $. Note that the phase generated by the RSO between the
$|\sigma \overline{\sigma} \rangle$ QDs configurations is the same
for both the RKKY and the Kondo-doublet (Eq.(\ref{twfm}) and
Eq.(\ref{bu2})), resulting in the $\cos{2 k_s R }$ terms in their
$\langle S_R.S_L\rangle$ correlations.

To distinguish between the Kondo-doublet and RKKY contributions
in the experimental setup one can rely on the different dependences
of their correlation energies on $k_F R$ . While the RKKY
depends on $2k_F R$, the Kondo-doublet depends just on $k_F R$,
see Fig. \ref{fig2}. This is due to the fact that
the fourth order RKKY path involves two band excitations, an
electron and a hole, whereas the second order Kondo-doublet
path involves just a hole. For the same reason the RKKY decays more quickly, as a
function of $R$, than the Kondo-doublet. The same lateral gate voltages that generate
the transversal RSO electric field can be used to modify $k_F$.

\textit{Conclusions}. We have found that the strength of the Kondo-doublet interaction is more than an order of magnitude greater than that of the RKKY in semiconductor Kondo QD-QW systems. Therefore, to assume that the interaction between Anderson-Kondo impurities is given by the RKKY process alone, as is usually done, misses the physics of the problem.  We also found that the wire Rashba spin orbit coupling can be used to control the spin correlation between
QDs through the effects that it induces on the
Kondo-doublet interaction. The proposed experimental setup allows
also to discriminate between Kondo-doublet and RKKY contributions.  Although we center the present work on the semiconductor QDs-QW data, the system can be realized also with other nano-techniques, as reported in Ref.\cite{agwire}. 

Finally, we briefly address the implications of this work on the Kondo screening cloud
pursuit \cite{hu,pascal2,pascal3}. The Kondo-doublet screening cloud is essentially the
single impurity cloud \cite{varma} amplified by the resonant ``pinning" of one of its components to both QDs. Thus, the measurement of the Kondo-doublet predictions is a direct test of the single impurity Kondo cloud.

 I acknowledge interesting conversations with G. Usaj, and thanks the CONICET
(Argentina) for partial financial support.

\bibliography{dkso}

\end{document}